\documentclass[pra,superscriptaddress,a4paper,twocolumn,showpacs]{revtex4-1}
\usepackage{graphicx, color, subfigure, amsmath}

\begin{document}

\title{A Photonic Implementation for the \\
Topological Cluster State Quantum Computer}

\author{David A. Herrera-Mart\'i}
\email{dherrera@imperial.ac.uk}
\affiliation{Institute for Mathematical Sciences, Imperial College London, London SW7 2BW, United Kingdom}
\author{Austin G. Fowler}
\affiliation{Centre for Quantum Computer Technology, University of Melbourne, Victoria, Australia}
\author{David Jennings}
\affiliation{Institute for Mathematical Sciences, Imperial College London, London SW7 2BW, United Kingdom}
\author{Terry Rudolph}
\affiliation{Institute for Mathematical Sciences, Imperial College London, London SW7 2BW, United Kingdom}

\begin{abstract}
A new implementation of the topological cluster state quantum computer is suggested, in which the basic elements are linear optics, measurements, and a two-dimensional array of quantum dots. This overcomes the need for non-linear devices to create a lattice of entangled photons. We give estimates of the minimum efficiencies needed for the detectors, fusion gates and quantum dots, from a numerical simulation.
\end{abstract}

\date{\today}

\maketitle

\section{Introduction}

The search for quantum systems tha are both resilient to errors and can be manipulated while preserving quantum coherence is arguably one of the most challenging research lines in the field quantum computation. Optical architectures are good candidates to build a quantum computer, because photon polarization is a quintessentially quantum two-level system, photons are cheap to produce and can be manipulated with an extrememly high degree of precision. The main problem with them is that it is difficult to make two photons interact, which precludes a generalized way of building two qubit gates. Several schemes for fault-tolerant computation with linear optics have been proposed \cite{knill2001scheme,dawson2005,varnava2007loss}. However, implementation of these schemes remains challenging due to their very high resource overheads and use of feed-forward processing.

Raussendorf and Briegel \cite{raussendorf2001one} introduced an appealing version of measurement-based quantum computation which uses a cluster state as a means for propagating the quantum correlations. Provided with this highly entangled state one can simulate any quantum gate, up to some correctable rotation, by just measuring single qubits in the cluster in different directions. It allows for highly parallelizable computations, since independent threads of computation can be performed concomitantly in separated areas of the cluster. Note that only measurements and classical postprocessing are necessary, since the cluster state, which carries the quantum correlations, is given prior to the computation. A recent development \cite{raussendorf2006fault, raussendorf2007topological} has been to combine this method of quantum computing with ideas regarding topological protection of quantum information, and in this paper we look at implementing these ideas in an optical setting.

There has been much work on analyzing the methods of the creation of optical cluster states \cite{nielsen2004optical, browne2005resource}. Considerable resource savings would be attained if we were able to create good sources of single photons which were already in a cluster state. We will focus here on one proposal for doing so, based on quantum dots \cite{lindner2009}. Related proposals for atomic systems can be found in \cite{nielsen-molmer2010, li2010demand}.

In particular we consider an architecture in which banks of the photonic cluster state machine gun sources of \cite{lindner2009} produce one-dimensional cluster states. These are fused \cite{browne2005resource} into a cluster state capable of the aforementioned topologically protected encodings, and then measurements are performed on the resulting cluster state to perform the desired computation. We compute fault-tolerant thresholds for this proposal. A primary concern, and one of the main motivations of the paper, is that the probabilistic nature of the fusion gates --- which necessitates multiple attempts at forming certain bonds in the cluster state --- could drastically lower the thresholds for fault-tolerance.

This paper is organized as follows. In section II both the topological cluster state computer and the cluster state machine gun source are briefly reviewed. In section III we will explain how to construct the three dimensional cluster state using these elements. In section IV, we will present and discuss our results.

\section{Review}

We now briefly review the scheme of quantum computation proposed by Raussendorf \emph{et al.~}\cite{Raus07,Raus07d}. We will also describe two devices introduced in \cite{browne2005resource} and \cite{lindner2009}, which naturally complement each other in creating the cluster state in three dimensions.

\subsection{The Topological Cluster State Quantum Computer}

Drawing from topological protection ideas \cite{Brav98,Denn02}, which attain very high thresholds \cite{Raus07,Fowl08,Wang09} but are currently rather far from physical realization \cite{Stoc08,devitt2009architectural,VanM09,DiVi09}, Raussendorf \emph{et al.} devised a measurement based model \cite{Raus07,Raus07d,Fowl09} which features advantages from topological codes. In their scheme a cluster state is used to build a three dimensional version of the surface code \cite{Brav98}. This three-dimensional variant will inherit the CSS structure of surface codes \footnote{This amounts to saying that X and Z errors can be considered independently.}, which can also be seen in the fact that it is defined in two dual lattices --- a property that will be useful to see how syndromes can be extracted by measuring only products of Pauli X stabilizers.

Since a surface code has trivial topology it cannot encode a qubit unless some degree of freedom is released. This can be done by relaxing (disregarding) one or more stabilizer constraints, or, in a more pictorial way, by creating holes in the surface to attain a topologically nontrivial shape. A suitable way of encoding qubits is to measure out two separate groups of stabilizers (from now on termed `holes' or `defects') effectively creating a two dimensional subspace in the code. A similar reasoning applies in one higher dimension, where one dimension can be singled out as time and the remaining two are seen as a surface in which the holes move.

\begin{figure}
  \centering
  \subfigure[]{
  \includegraphics[scale=.18]{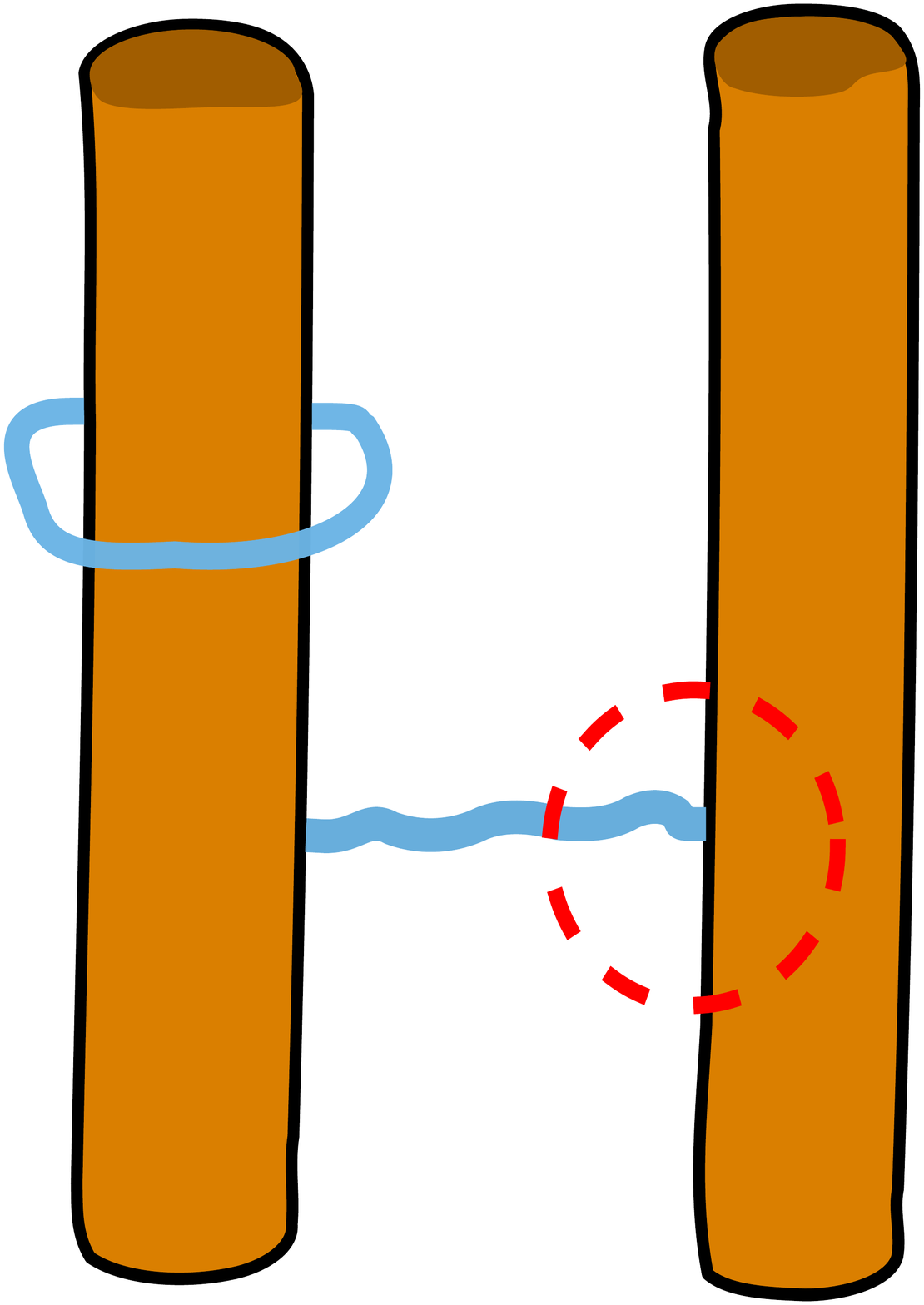}}
  \subfigure[]{
  \includegraphics[scale=.18]{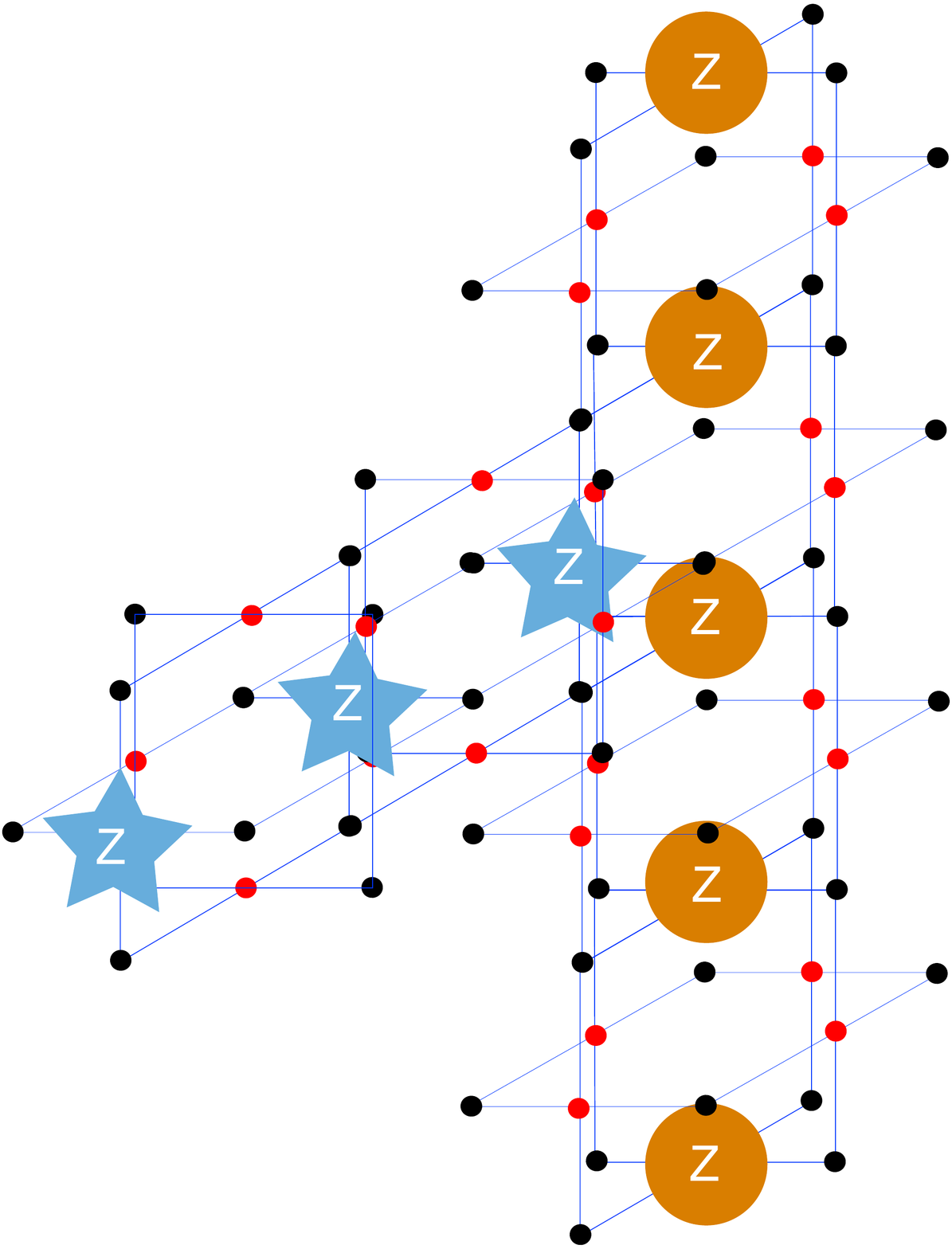}}
  \caption{We consider only the primal lattice, focusing only on one error channel, namely Z errors. X errors will be dealt with looking at the dual lattice. (a) Macroscopic view. Defects (holes) are measured in the Z basis, whereas the rest of the qubits in the cluster will be measured in the X basis. (b) Microscopic view. The stars represent a chain of phase flip operators that goes from one defect to another. This chain doesn't belong to the stabilizer group but still commutes with all its elements, so it will act non-trivially on the qubit encoded in the defects. The same happens with the chain that winds around one of the defects, i.e. it will change the logical state of the encoded qubit.}
  \label{fig_macmic}
\end{figure}

\begin{figure}
  \includegraphics[scale=.2]{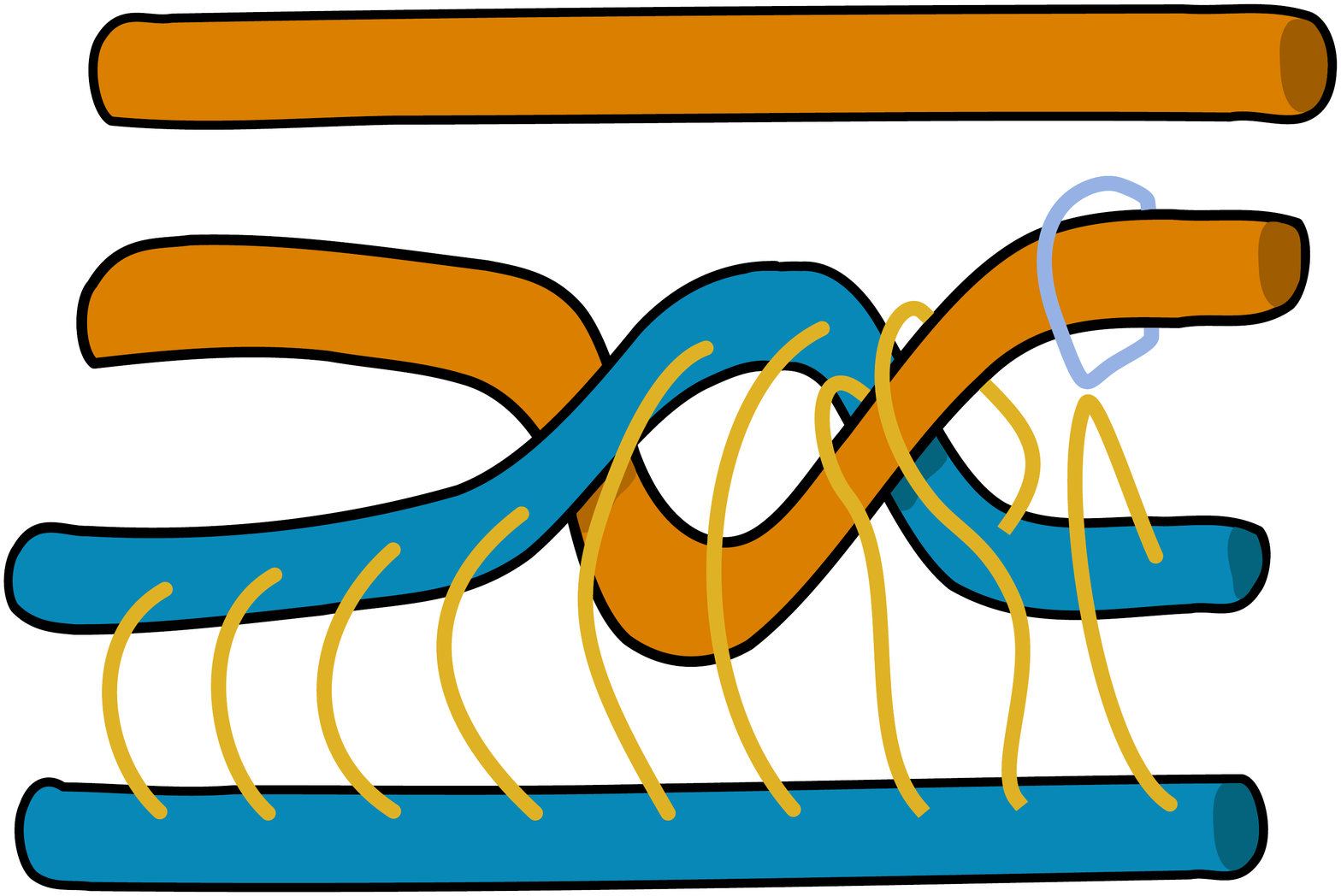}
  \caption{CNOT gate between a primal and a dual logical qubit, as a result of braiding the respective defects. The lighter lines represent a correlation surface compatible with the Z measurements pattern. Whatever information was encoded in the dual defects will be passed on to the primal defects as a result of the X measurements.}
  \label{fig_braid}
\end{figure}

\begin{figure}
  \subfigure[]{
  \includegraphics[scale=.13]{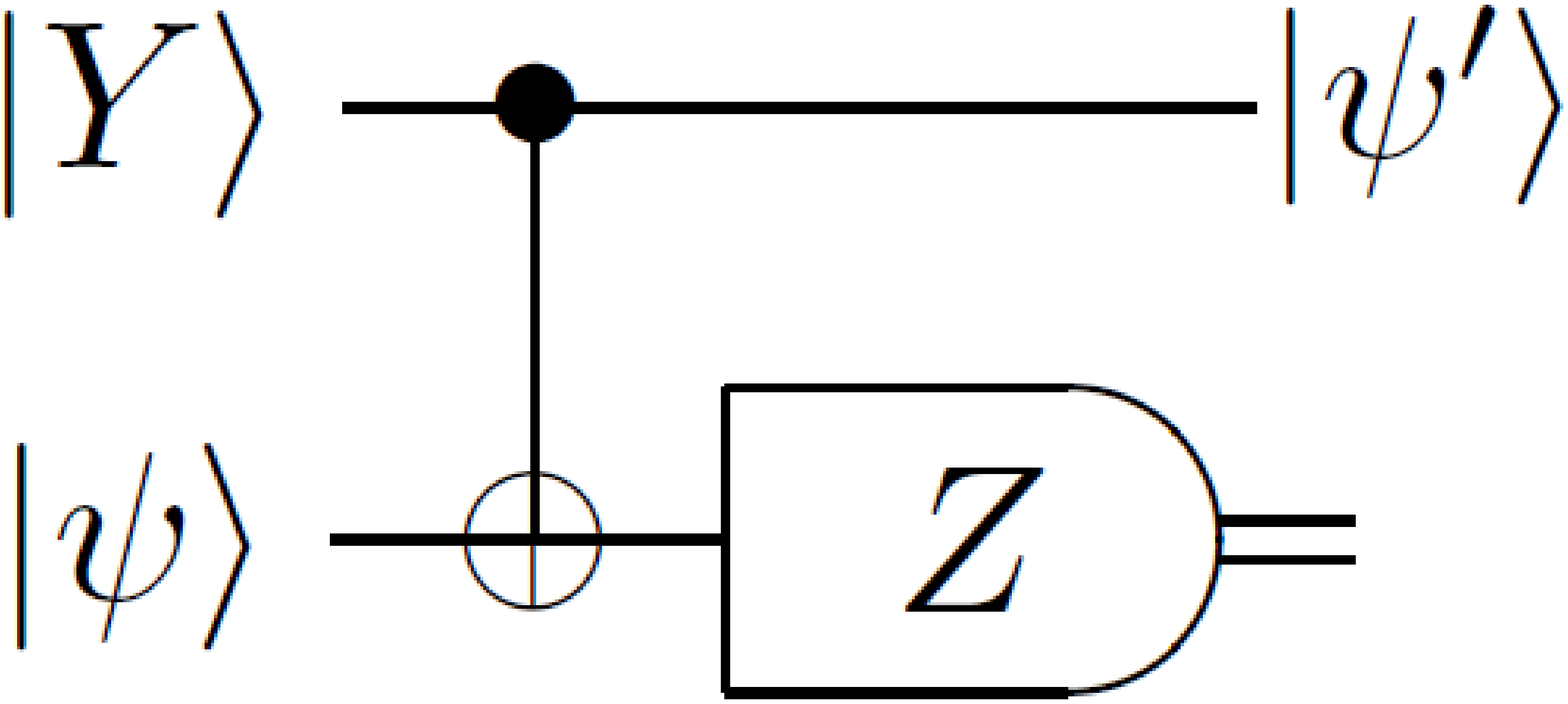}}
  \subfigure[]{
  \includegraphics[scale=.15]{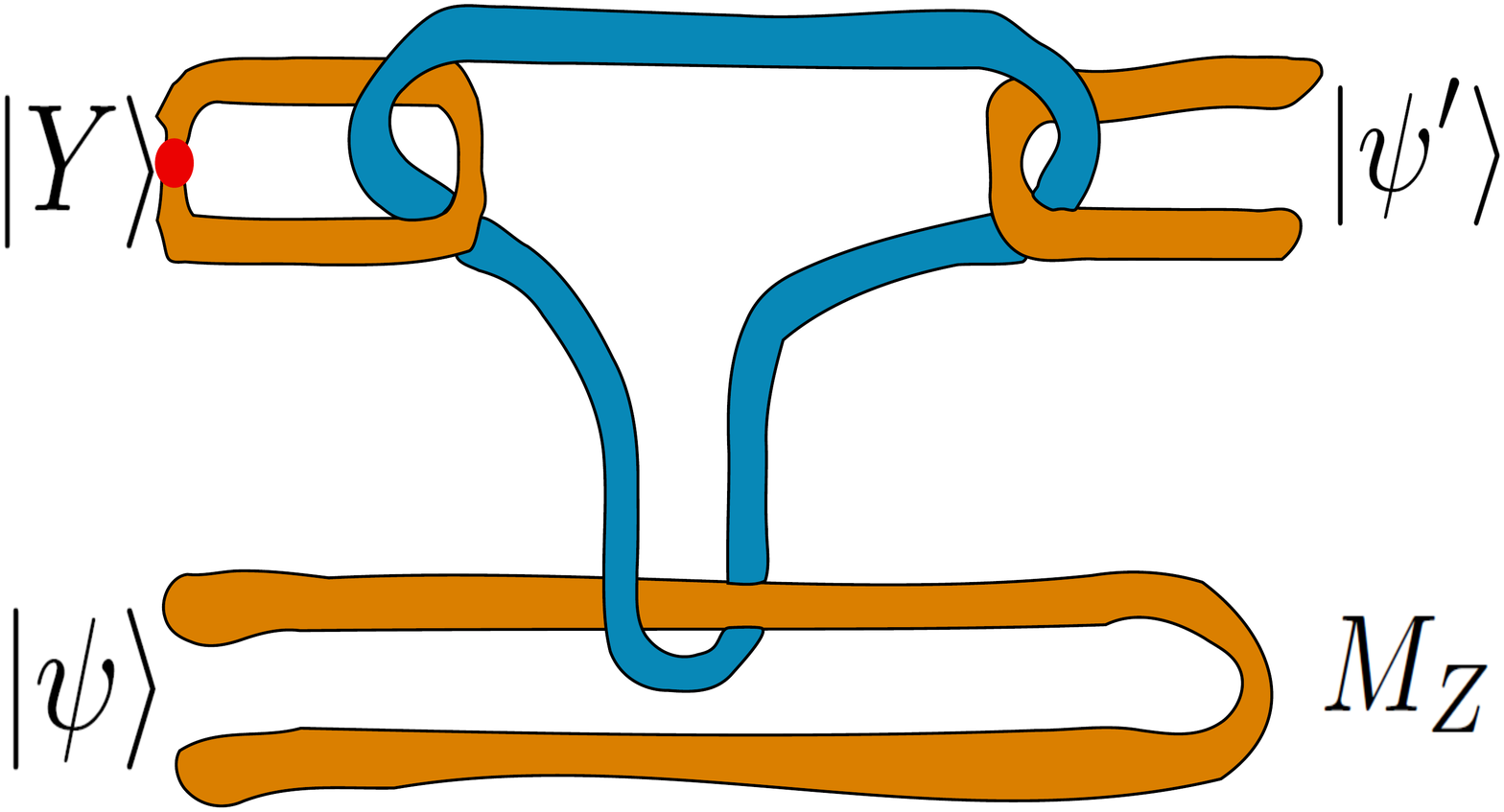}}
  \caption{(a) Gate teleportation circuit. A probabilistic phase gate K. Given the \emph{magic state} $|Y\rangle = \frac{1}{\sqrt2}|0\rangle + i|1\rangle$, the gate teleportation circuit will return $|\psi'\rangle = X^{M_z}K|\psi\rangle$, where $M_Z \in \{0,1\}$. Similar results hold for $\frac{\pi}{8}$ gates. (b) Topological gate teleportation circuit.}
  \label{fig_phase}
\end{figure}

The stabilizer structure of the cluster state will ensure that an X error occurring on the primal lattice is equivalent to a number of Z errors occurring in the dual lattice, and viceversa. This allows us to consider only Z errors, which will flip the stabilizers of the two adjacent cells that share the faulty qubit, these stabilizers are of the form $S_{cell}=\Pi^6_{i=1} X_i$, where we have an X for each face of the cell. One of the beautiful features of this scheme is that measuring the code qubits in the X direction will at the same time give us the syndrome, and propagate the code forwards in time \cite{Raus07d}.

As mentioned, the cluster state is defined in two interlaced lattices. This is best understood in the microscopic picture so we will not dwell much upon it. However it is important to keep in mind that there exist two types of hole (or defect), \emph{primal} and \emph{dual}, depending upon the (sub-)lattice in which they evolve. It is useful to keep in mind that primal and dual defects cannot touch as they each live in a separate lattice. However, the braiding of primal and dual holes along the computational direction will result in a set of \emph{correlation surfaces} (sheets of stabilizers which carry the correlations from one time-slice of the computation to another one) which are compatible with the topology of the cluster. The correlation surfaces compatible with the braiding in figure \ref{fig_braid} can be shown to enact a CNOT gate between defects of different kind. We also need to have a CNOT gate between two defects of the same kind, and this is also possible --- a method for doing so is used in the circuit of fig.~\ref{fig_phase}.

In order to achieve universal quantum computation the set of gates has to be expanded beyond CNOT and preparations and measurements in the Pauli group. The way this is done is by distilling \emph{magic states} \cite{bravyi2005universal}, which can be used to attempt phase and $\pi/8$ gates via topologically protected circuits (see fig.~\ref{fig_phase}). In order to simulate Hadamard gates, we need three such gate teleportation circuits, thus completing the set of universal gates.

\subsection{The Photonic Cluster State Machine Gun}

As its name suggests, the photonic cluster state machine gun fires a stream of polarized photons in a linear cluster state \cite{lindner2009}. Although it is in principle possible to build such a device using different quantum systems, we will focus on quantum dots as their spontaneous emission rate is of order of picoseconds, the allowed transitions are better separated than in atoms, and they can be tailored in the lab to fine tune certain parameters such as the frequency of the optical transitions \cite{imamog1999quantum,imamoglu2003quantum,loss2000electron}.

\begin{figure}
  \centering
  \subfigure[]{
  \includegraphics[scale=.2]{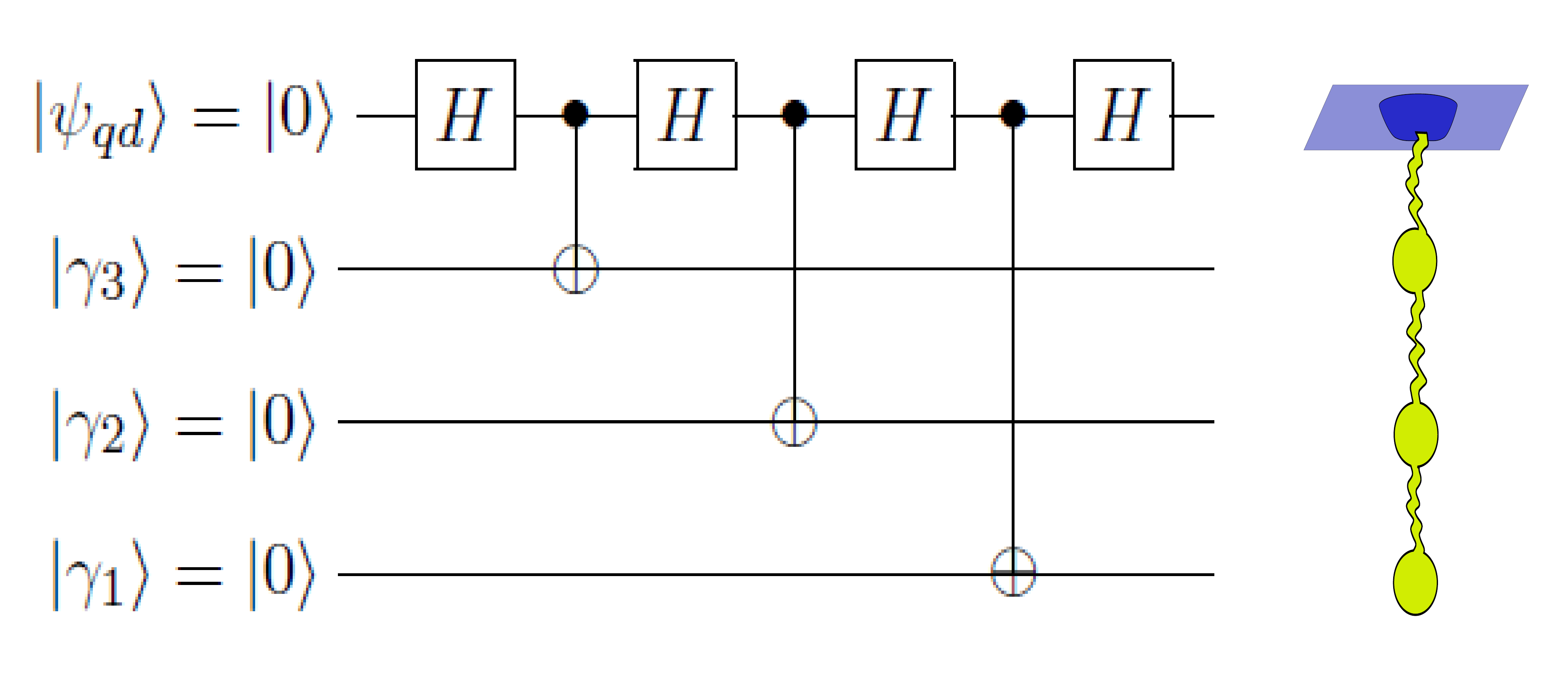}}
  \subfigure[]{
  \includegraphics[scale=.21]{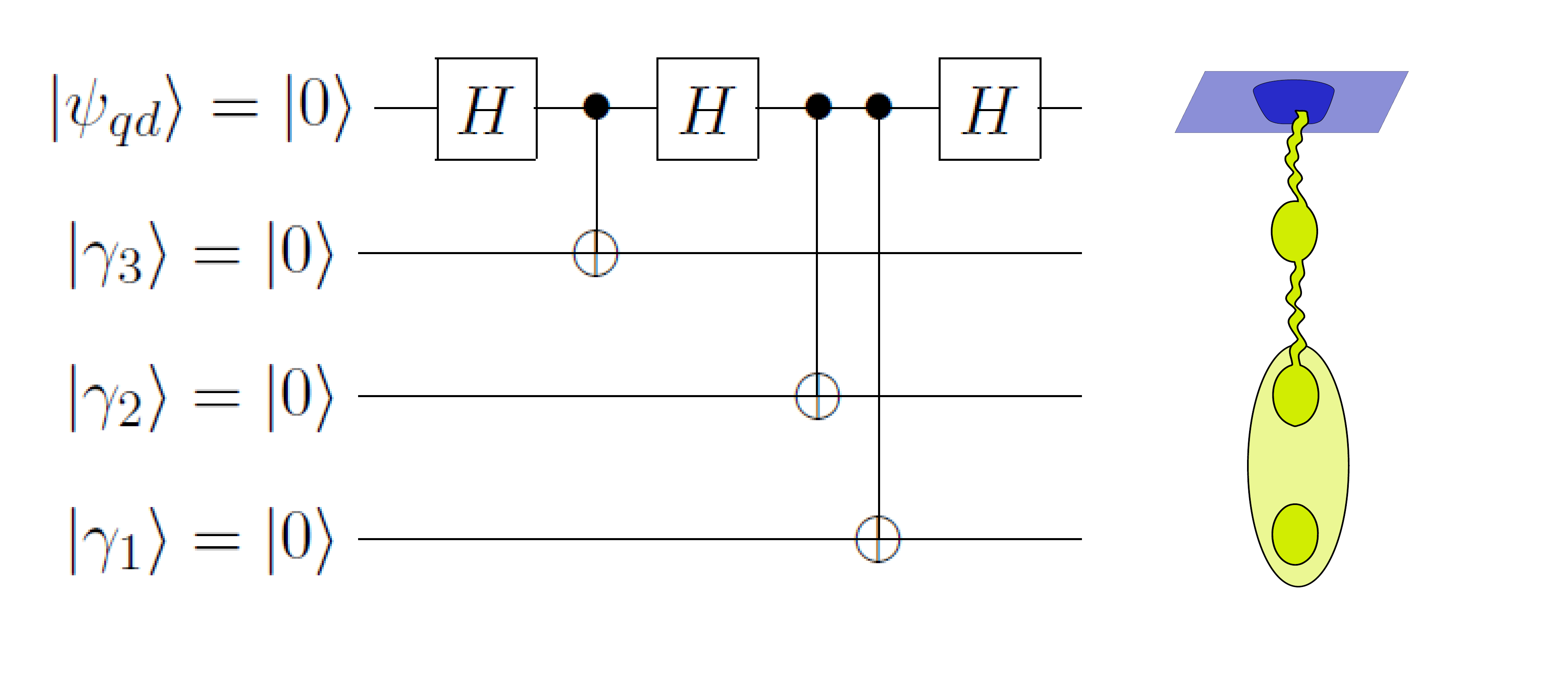}}
  \caption{Equivalent circuit to the precession-emission process. Although it offers valuable insight, this circuit obviously doesn't encompass the physics inside the dot. (a) Three qubit linear cluster state. With a suitable choice of logical states, i.e. $|0\rangle \equiv |R\rangle$ and $|1\rangle \equiv -|L\rangle$, one can identify the rotation $R_{Y}(\frac{\pi}{2}) = e^{-iY\frac{\pi}{4}}$ with a Hadamard gate. Emission of a photon will be represented by a CNOT gate, since the photon's polarization will depend on the spin of the quantum dot. (b) Two qubits linear cluster state. One of them is a RE qubit. If the quantum dots are left to precess enough time to describe a rotation $R_{Y}(4\pi) \equiv I$, no gate will operate on the dot, which can emit qubits in a relative GHZ state.}
  \label{fig_mg}
\end{figure}

At the risk of oversimplifying, a quantum dot can be described as a two degenerate levels system consisting of the ground states $\{|\uparrow\rangle, |\downarrow\rangle\}$ and the excited states $\{|\Uparrow\rangle, |\Downarrow\rangle\}$. These represent the quantum dot with an electron in the conduction band, and the quantum dot with two electrons in the conduction band and one hole in the valence band (called \emph{trion}), respectively. The selection rules ensure that only the transitions $|\uparrow\rangle\leftrightarrow|\Uparrow\rangle$ and $|\downarrow\rangle \leftrightarrow|\Downarrow\rangle$ will occur, where the decays will be followed by emission of circularly polarized light. If we introduce a magnetic field transversal to the spin direction, say in the Y direction, the electron will precess with frequency $\omega_P = g_e \mu B_Y/\hbar$. With a suitable choice of logical states, we can identify the rotation $R_{Y}(\frac{\pi}{2}) = e^{-iY\frac{\pi}{4}}$ with a Hadamard gate. The precession will implement this gate every $T_H = \pi/2\omega_P$ seconds. If at these intervals linear polarized light is shined onto the dot, a coherent superposition of trion states will be created, which will decay almost immediately. This is enough to create a linear cluster state:

\begin{eqnarray}\label{ex_pmg}\nonumber
|\uparrow\rangle &\stackrel{H}{\rightarrow}& |\uparrow\rangle + |\downarrow\rangle\stackrel{E}{\rightarrow} |\uparrow\rangle|R\rangle + |\downarrow\rangle|L\rangle \\\nonumber
&\stackrel{H}{\rightarrow}&(|\uparrow\rangle + |\downarrow\rangle)|R\rangle + (-|\uparrow\rangle + |\downarrow\rangle)|L\rangle\\\nonumber
&\stackrel{E}{\rightarrow}&(|\uparrow\rangle|R\rangle + |\downarrow\rangle|L\rangle)|R\rangle + (-|\uparrow\rangle|R\rangle + |\downarrow\rangle|L\rangle)|L\rangle\\\nonumber
&\stackrel{H}{\rightarrow}& |\uparrow RR\rangle + |\downarrow RR\rangle - |\uparrow LR\rangle + |\downarrow LR\rangle\\
&-& |\uparrow RL\rangle - |\downarrow RL\rangle - |\uparrow RL\rangle + |\downarrow LL\rangle
\end{eqnarray}

If we identify $|0\rangle \equiv |R\rangle$ and $|1\rangle \equiv -|L\rangle$, then expression \ref{ex_pmg} can be seen to represent a three qubit cluster state.

Figure \ref{fig_mg} depicts the generation of a cluster state consisting of 3 photons. The cluster creation in fig.~\ref{fig_mg}(a) is best seen within the stabilizer description:
\begin{eqnarray}\nonumber
&&\begin{vmatrix}
    Z & I & I & I \\
    I & Z & I & I \\
    I & I & Z & I \\
    I & I & I & Z
\end{vmatrix}\stackrel{1}{\Rightarrow}
\begin{vmatrix}
    X & X & I & I \\
    Z & Z & I & I \\
    I & I & Z & I \\
    I & I & I & Z
\end{vmatrix}\stackrel{2}{\Rightarrow}
\begin{vmatrix}
    Z & X & I & I \\
    X & Z & X & I \\
    Z & I & Z & I \\
    I & I & I & Z
\end{vmatrix}\stackrel{3}{\Rightarrow}\\&&
\begin{vmatrix}
    X & X & I & X \\
    Z & Z & X & I \\
    X & I & Z & X \\
    Z & I & I & Z
\end{vmatrix}\stackrel{4}{\Rightarrow}
\begin{vmatrix}
    Z & X & I & X \\
    X & Z & X & I \\
    Z & I & Z & X \\
    X & I & I & Z
\end{vmatrix}\stackrel{5}{\Rightarrow}
\begin{vmatrix}
    \textbf{X} & \textbf{Z} & \textbf{I} \\
    \textbf{Z} & \textbf{X} & \textbf{Z} \\
    \textbf{I} & \textbf{Z} & \textbf{X} \\
\end{vmatrix}
\end{eqnarray}

Here, each step corresponds to a Hadamard gate applied to the quantum dot followed by a CNOT gate, except the steps 4 and 5, which represent a single Hadamard and measuring out the quantum dot, respectively. One of the most appealing features of the machine gun is that Pauli errors in the quantum dot (due to dephasing etc.) amount to a local error in the stream of photons, which can be shown by manipulating its stabilizers.

In our scheme we will need to encode a qubit into several \emph{redundantly encoded} (RE) photons. This is very easy to attain with the machine gun: one only has to let the quantum dot precess around the Y axis for a time corresponding to a $4\pi$ radians rotation before exciting it\footnote{We can equally consider a rotation of $2\pi$, since $R_{Y}(2\pi) \equiv -I$, since this can also be used to obtain RE qubits up to phases. }. Then it is easy to see that no Hadamard will act on the dot, so the new photon will be in a RE state with the previous photon(s) --- see figure \ref{fig_mg}(b).

\subsection{The Fusion Gate}

\begin{figure}
  \subfigure[]{
  \includegraphics[scale=.17]{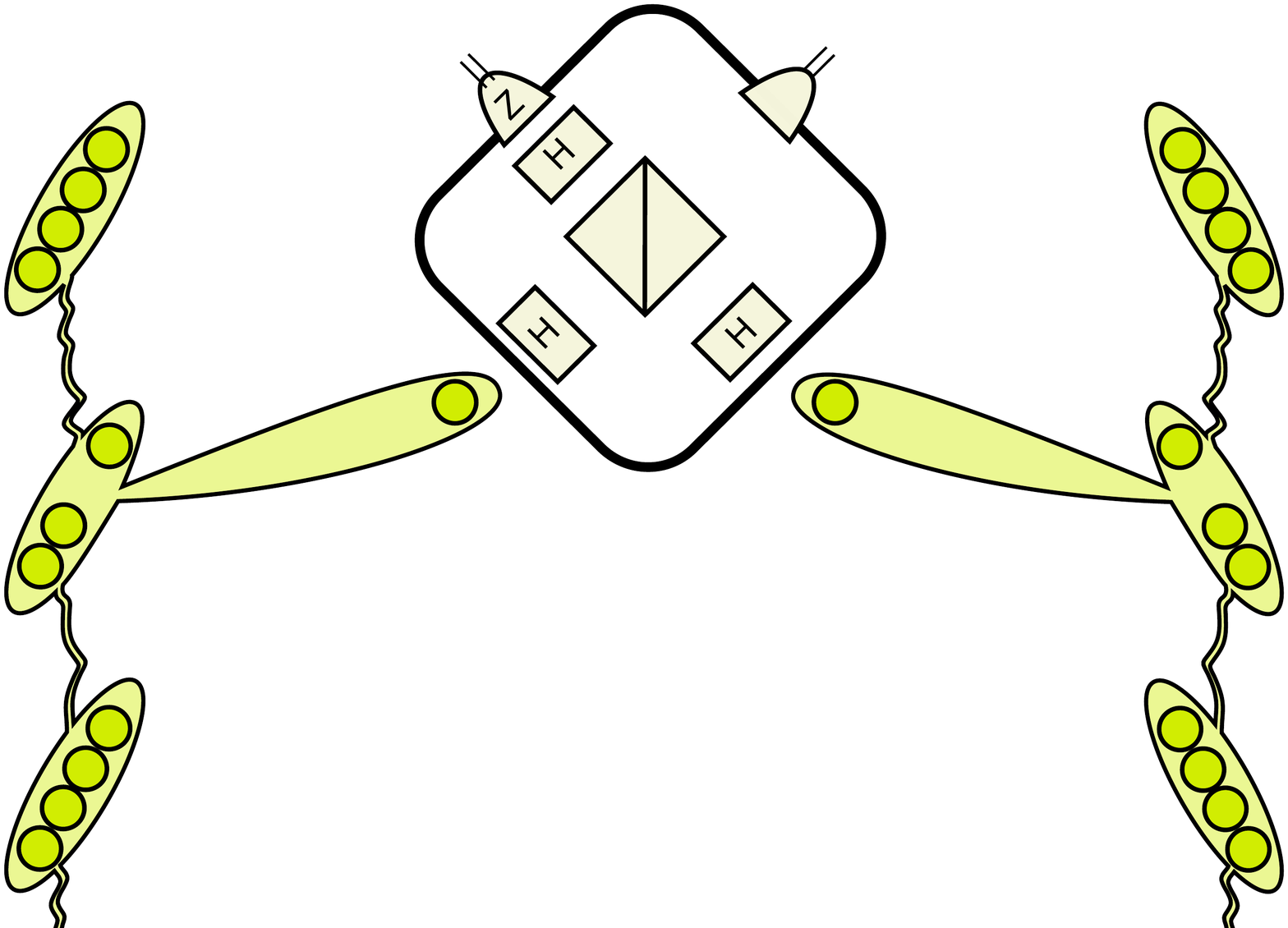}}
  \subfigure[]{
  \includegraphics[scale=.13]{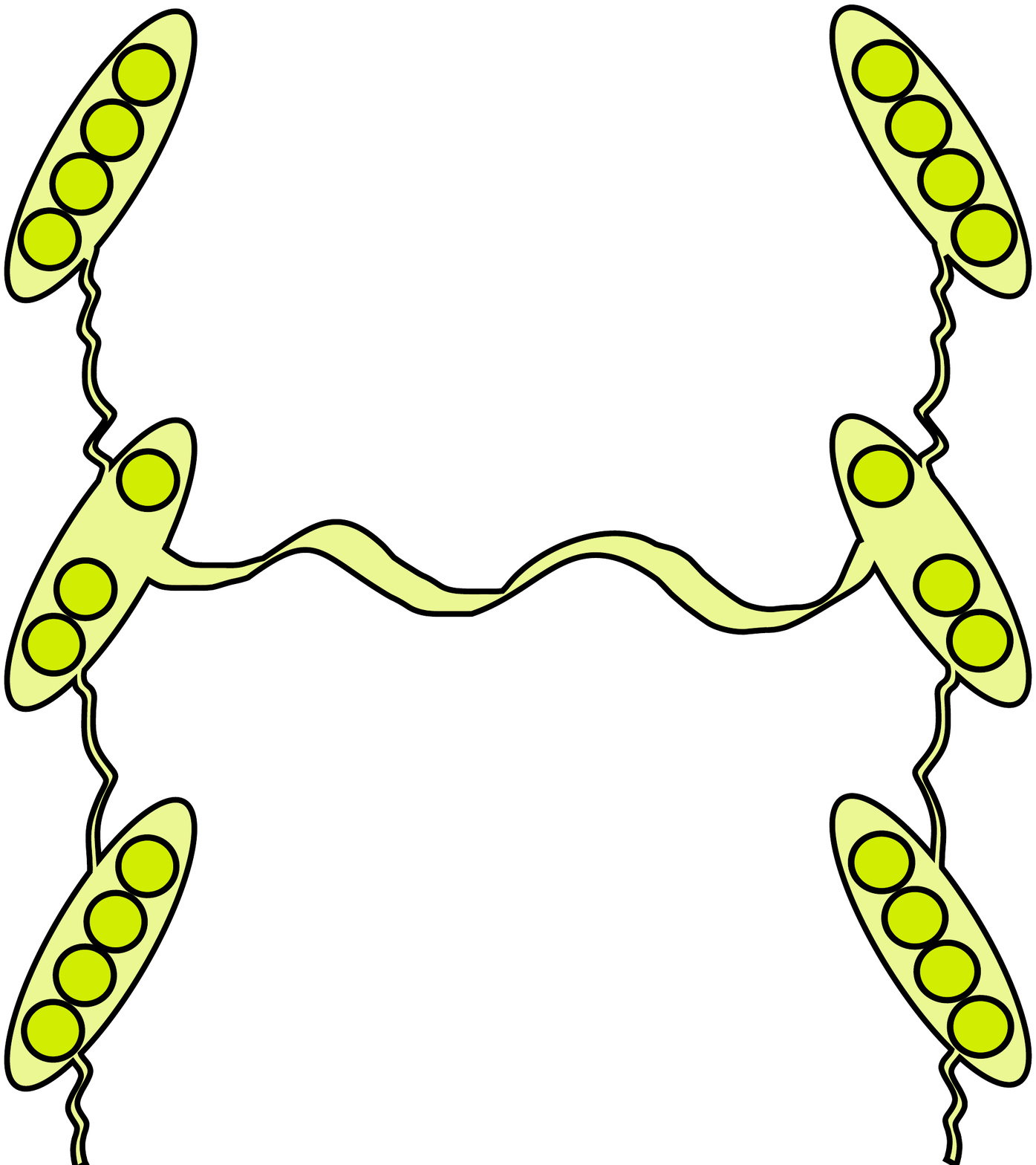}}
  \caption{(a) One photon from each stream is used to attempt a fusion gate. (b) Upon success, a link between the streams will be created. If this is done in all directions it will give rise to an arbitrary large, albeit incomplete, three dimensional cluster state.}
  \label{fig_fusion}
\end{figure}

The gate we present here is a slight variation of the ones proposed in \cite{browne2005resource}, since here we need to create links between RE qubits rather than simply fuse them into larger RE sets. The gate is depicted in fig.~\ref{fig_fusion}.

The gate will first apply a Hadamard gate to the photons, pushing them out from the RE qubit (making them stick out like a leafy branch as it were). The next part of the gate is equivalent to a Type-I fusion \cite{browne2005resource}. If the fusion fails it will effectively collapse the qubits in the Z basis, and that is why we need to differentiate the input photons from the rest of the RE qubit, since such failure would then destroy coherence in the RE qubit. If it is successful, there will be two RE qubits linked to a third photon. In order to create a link between the two extremal RE qubits, one has to measure out the middle photon in the Y basis. This will give a cluster state, up to a phase gate K, which satisfies $KYK^{\dagger} = X$, $KZK^{\dagger} = Z$ and can be dealt with in the measurement stage.

A simplified version of how our fusion gate operates can be described considering two Bell pairs. In terms of stabilizers, the evolution exposed above takes the form:

\begin{equation}
  \begin{vmatrix}
    X_1 & X_2 \\
    Z_1 & Z_2 \\
  \end{vmatrix}
  \begin{vmatrix}
    X_3 & X_4 \\
    Z_3 & Z_4 \\
  \end{vmatrix}\stackrel{T-I_3}{\Longrightarrow}
  \begin{vmatrix}
    X_1 & Z_2 & I_4\\
    Z_1 & X_2 & Z_4\\
    I_1 & Z_2 & X_4\\
  \end{vmatrix}\stackrel{M^Y_2}{\Longrightarrow}
  \begin{vmatrix}
    \mathbf{Y_1} & \mathbf{Z_4} \\
    \mathbf{Z_1} & \mathbf{Y_4} \\
  \end{vmatrix}
\end{equation}

The fusion gate succeeds half of the time. One possible concern is what happens when the detectors inside the fusion gate are not perfect. It can be shown that even if one or both photons are lost, the resulting state still has a (smaller) probability of being in a cluster state. However this is not very useful since we have no means of knowing whether the entanglement is there or not, and so we condition only on a successful detection of both photons.

\section{Building up the Cluster}

In this section we will present the main idea of the paper, namely how to use the concepts explained in previous section to build an optical quantum computer.

\begin{figure}
  \includegraphics[scale=.23]{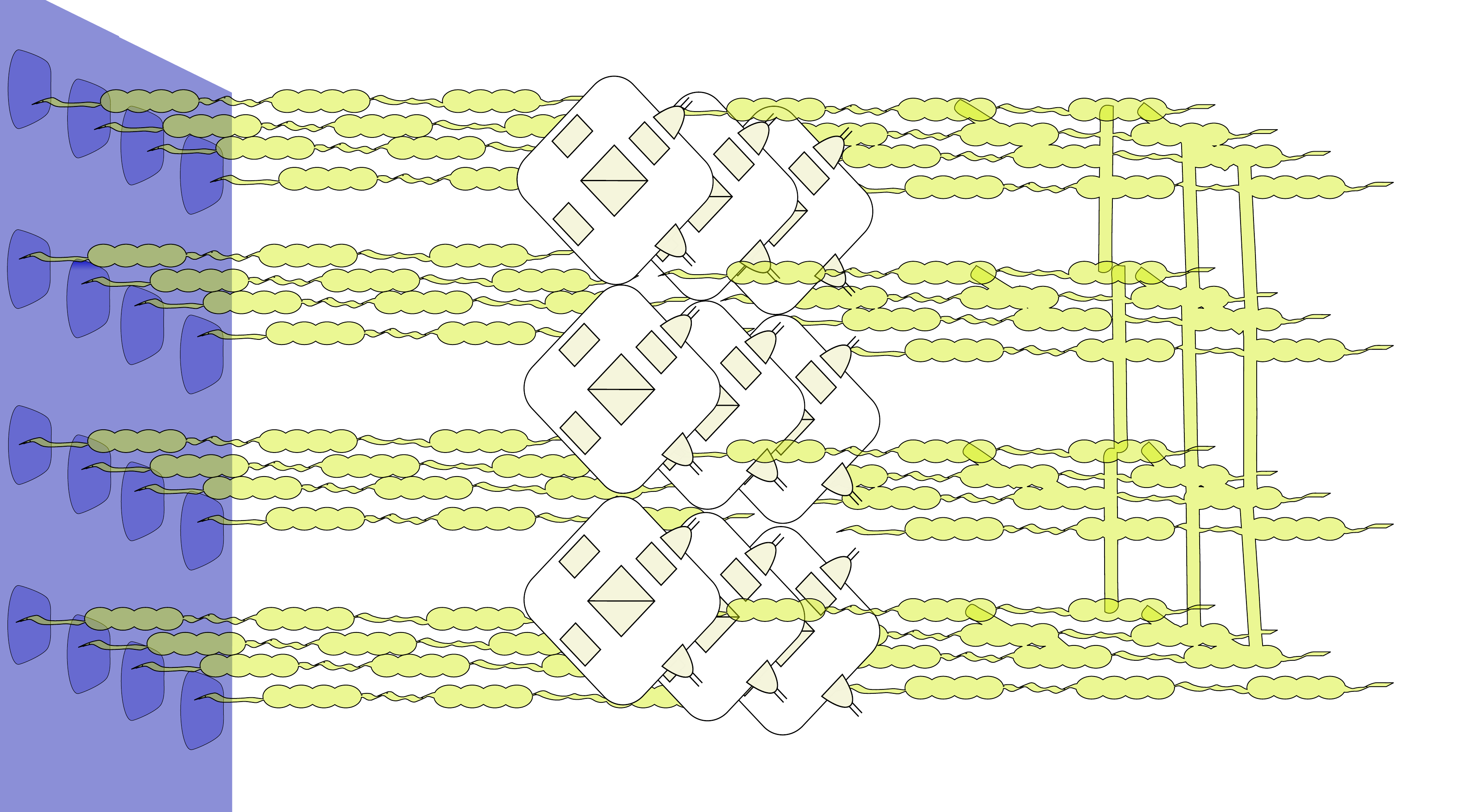}
  \caption{All elements considered so far assembled into a global view of our proposal. Fusion and detection stages should be connected to classical computers which can communicate in order to deal with the loss due to probabilistic gates.}
  \label{fig_global}
\end{figure}

In fig.~\ref{fig_global} we depict the basic idea of our proposal. We consider a two-dimensional array of quantum dots that behave as photonic machine guns firing parallel streams of photons. At a latter time, fusion gates probabilistically create links between the beams in such a way that when the photons leave this stage they will be with high probability in the cluster state described by Raussendorf (see fig.~\ref{fig_global}). Finally, another array of detectors will measure the incoming photons in the bases dictated by the desired quantum algorithm, creating a non-trivial topology in the cluster. The last two stages must be controlled by classical computers, so that heralded loss at fusion can be preprocessed and accounted for as well as to undo the possible Pauli errors that the syndrome may have unveiled.

It is important to notice that we need to use RE photons to maximize the probability of link creation, which goes as $p_{l} = 1 - 2^{-R}$, where R is the number of times we attempt a fusion, and to perform fusions in different directions. Depending on where in the cluster a qubit sits, i.e. in how many directions we have to try to fuse it with its neighbors, we will use a different number of RE photons. There are four types of qubit: qubits which have to be fused in the left-right direction, or in top-down direction will be composed of $2R+1$ RE photons, qubits which have to be fused in both directions will consist of $4R+1$ RE photons, and qubits which have to be measured out will only consist of one photon. It is easy to see that this doesn't pose any problem for synchronizing fusions. On average, a qubit will consist of $2R+1$ RE photons, which will be either used to create links or measured out in the process, leaving a final qubit consisting uniquely of one photon.

This is in many ways similar to the proposal by \cite{devitt2009architectural}, however it overcomes the need of building a highly efficient non-linear device known as a \emph{photonic module}. These modules are replaced with fusion gates, greatly reducing the difficulty of coupling photons at the cost of increasing the number of photons and decreasing the loss tolerance.

\subsection{Error Model}

We consider two basic types of error: computational errors and photon loss. Computational errors are modeled by one and two qubit depolarizing noise. Due to the finite lifetime of the trions, the quantum dots might precess longer than expected before emitting the photon. This results in faulty Hadamard gates with probability of depolarizing error $p_1$. Imperfections in the exciting pulse will result in further dephasing of the electron in the quantum dot just before emission. This is modeled by faulty CNOT gates in the equivalent circuit of fig.~\ref{fig_mg}, which introduce correlated depolarizing noise between two qubits with probability $p_2$. A more detailed discussion can be found in \cite{lindner2009}. A successful fusion will introduce correlated depolarizing noise with probability $p'_2$. Measurements are also allowed to be noisy, again with probability $p_1$.

Loss errors are themselves of two types. Loss at emission will happen with probability $p_{dot}$, whereas loss at detection will happen with $p_{det}$ --- this includes faulty detection in the fusion stage. For simulation purposes we identify $p_1=p_2=p'_2 \equiv p_C$ and $p_{dot} = p_{det} \equiv p_L$. This will enable us to find a threshold described by a curve varying only two parameters, in a way similar to \cite{stace2009thresholds,stace2010error}. It is however important to keep in mind that our error model is specific to the optical setting that we have proposed.

\section{Numerical Simulation}

We calculated threshold estimates for computational error probability as well as for loss probability. Here we present them and provide a numerically obtained tradeoff curve between the two.

Not only does the cluster state constitute the substrate for the computation: it also provides a code for error correction. An typical error will be a chain of phase flip operators. It will be detected upon measuring wrong sign stabilizer elements at its endpoints. The error correction procedure will then be to apply a series of phase flip operators in such a way that the resulting chain belongs to the trivial homology class, which is equivalent to saying that it belongs to the stabilizer group. Alternatively, error chains that end at the boundaries of the cluster or in a defect, or wind around a defect, will not be detected since they won't flip any stabilizer. For simulation purposes it is convenient to work with a cluster with no boundary, in which case a chain of errors winding around any of the dimensions of the 3-torus will leave no syndrome.

We used the standard method for estimating the threshold, namely carrying out a large number of montecarlo simulations for the error correction procedure and sampling them at different values of the error rates. This is tractable within reasonable time only for small cluster states, i.e.~clusters with code distance $d \leq 15$\footnote{The simulations we present here were done using the Imperial College High Performance Cluster. Preliminary estimates (100 simulations per point) were taken running our algorithm on an Intel Q8200 processor at 2.33 GHz and 3.5 Gb of RAM, for up to $d\leq13$.}. The basic idea is that, for error rates below the threshold, increasing $d$ will reduce the failure probability, since the error chain resulting from error correction will most likely belong to the trivial homology class, whereas for error rates above the threshold, going to higher $d$ will actually cause more errors than it can correct, in the sense that non-trivial chains will be present with high probability. In the limit of $d\rightarrow \infty$, the failure probability $P_F$ should look like:

\begin{equation}
\lim_{d\rightarrow \infty} P_F(p_C) = \frac{7}{8} H (p_C - p_{Th})
\label{step_function}
\end{equation}

where H is the step function and $p_{Th}$ is the threshold probability. The factor $\frac{7}{8}$ comes from the fact that one of the eight homology classes of the 3-torus is trivial.

A comment is in order about the parameter R in our algorithm. One might be tempted to increase R in order to get a complete cluster state with high probability. This turns out to be a bad idea for a series of reasons. First, as we said before, we need qubits with $2R + 1$ RE photons on average. This means that, even if fusions were successful 100\% of the times and $R=1$, we would still have on average 3 photons per qubit, which increases the effective error probability per qubit --- we confirmed this by getting a computational error threshold approximately a factor of six lower than $0.7\%$ \cite{Raus07d}, for no loss. Also, since RE photons form a relative GHZ state, the loss of one of them will cause the rest to completely lose coherence and collapse to a maximally mixed state, which renders the qubit useless.

The simulations were carried for $R=7$. We found that there is no benefit in setting $R\geq8$. This probably finds an explanation in the facts exposed above, since for $R\geq8$, $2R+1\geq17$ and a loss probability of 6\% will spoil the whole cluster.

\subsection{Thresholds}

Here we present the main result of the paper, a trade-off curve for the loss and computational error thresholds.

\begin{figure}
  \subfigure[]{
  \includegraphics[scale=.22]{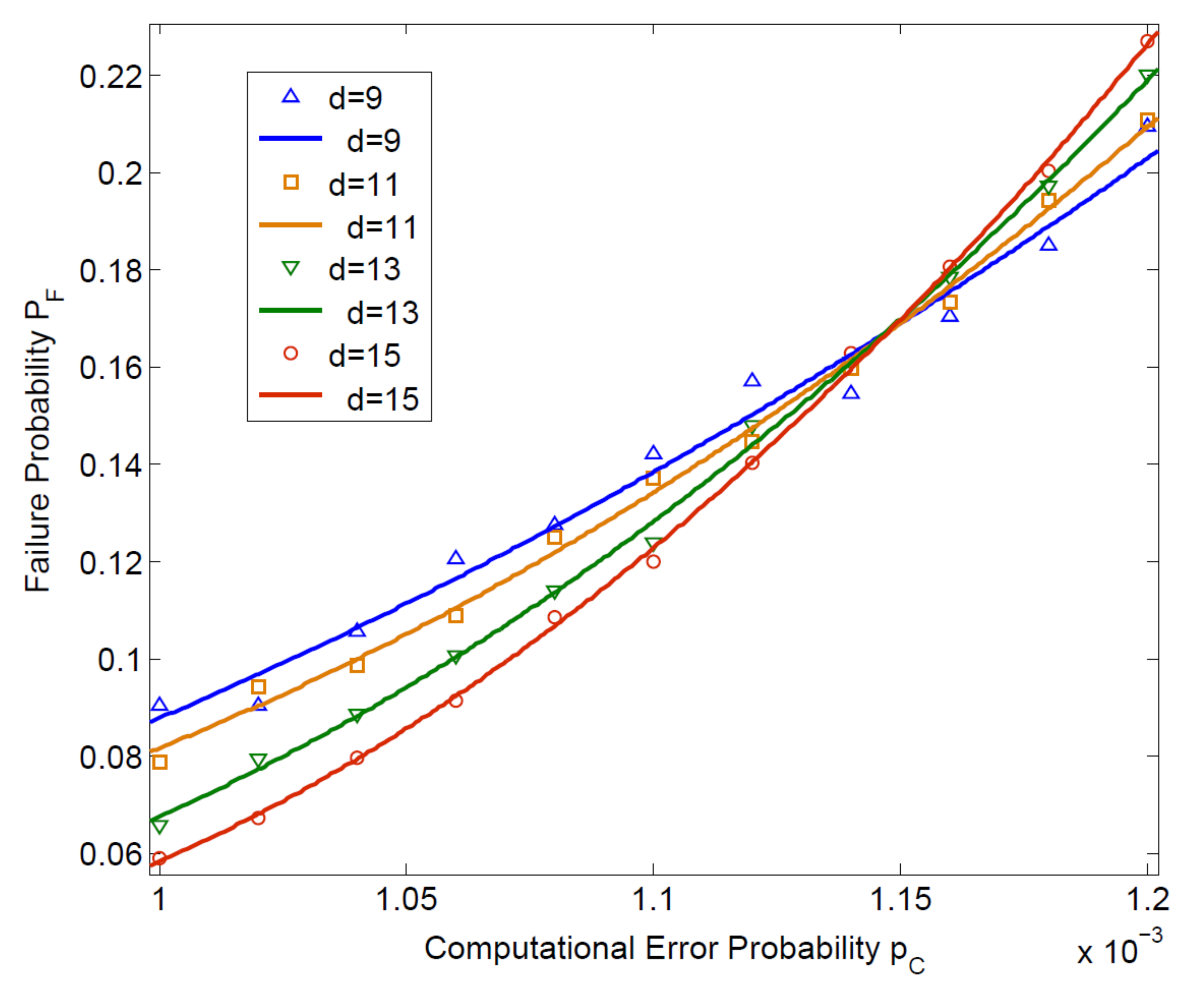}}
  \subfigure[]{
  \includegraphics[scale=.22]{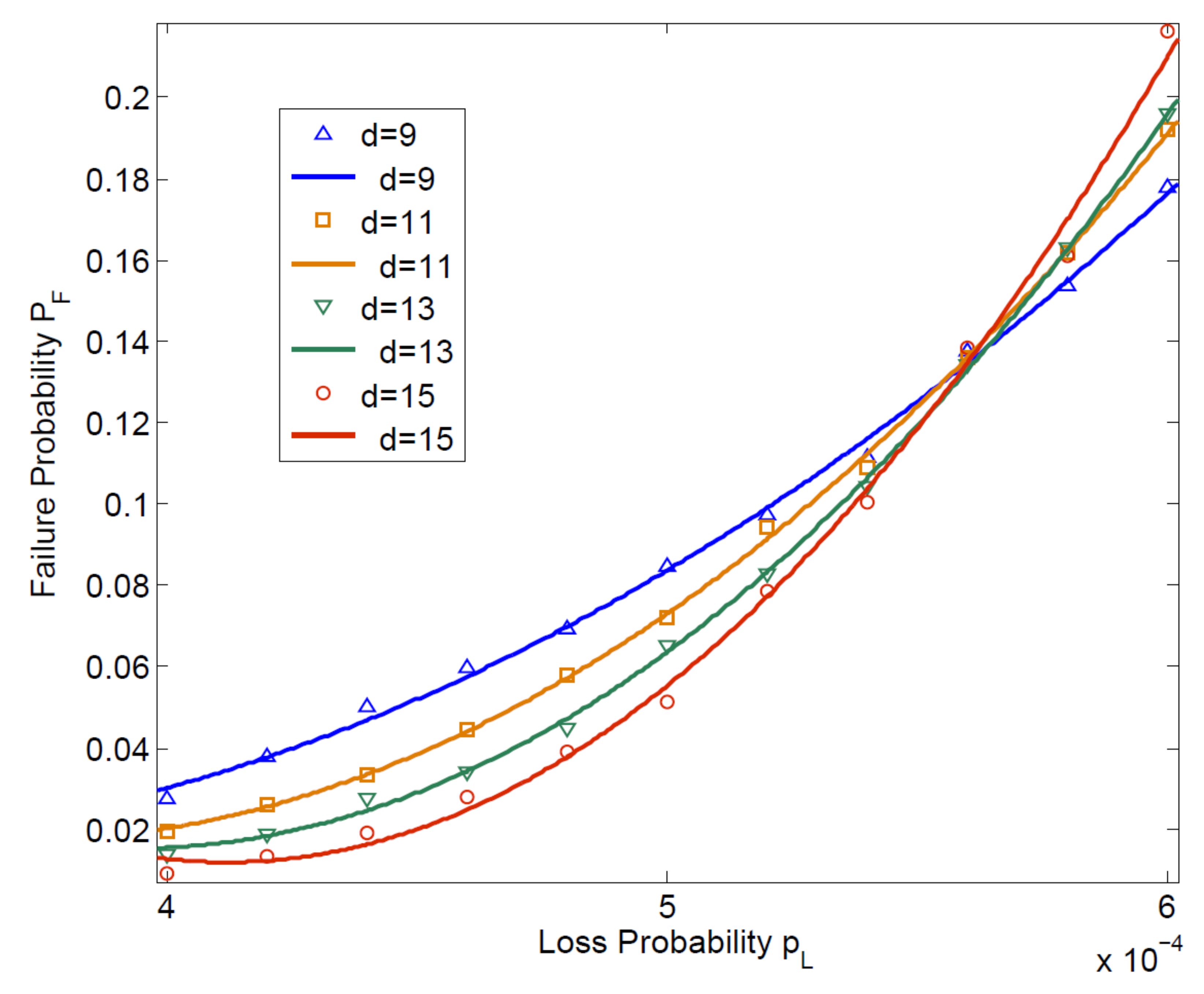}}
  \caption{Data for $d\leq7$ has not been included as finite size effects become too large at such low code distances.~(a) Fit for computational error only. (b) Fit for loss error only. Each point corresponds to the average of $10^4$ simulations (error bars not present). Crossing points of the curves for different d's, denoting the existence of a threshold, are observed at $p_C\approx1.14\times10^{-3}$ and at $p_L\approx5.3\times10^{-4}$.}
  \label{fit_errors}
\end{figure}

For no loss, we found that the threshold is $0.114\%$, only about 6 times smaller than in \cite{Raus07d} (see fig.~\ref{fit_errors}(a)). This is not surprising, since we are using several RE photons to encode a qubit, so the effective error rate per qubit will be necessarily higher than the error rate per photon.

The threshold for loss with no computational error is $0.053\%$ (see fig.~\ref{fit_errors}(b)). This is, however, unrealistically small. Our qubits are very sensitive to photon loss. An encoding less naive than mere RE photons \cite{varnava2007loss} would perhaps help to improve the error.

\begin{figure}
  \includegraphics[scale=.22]{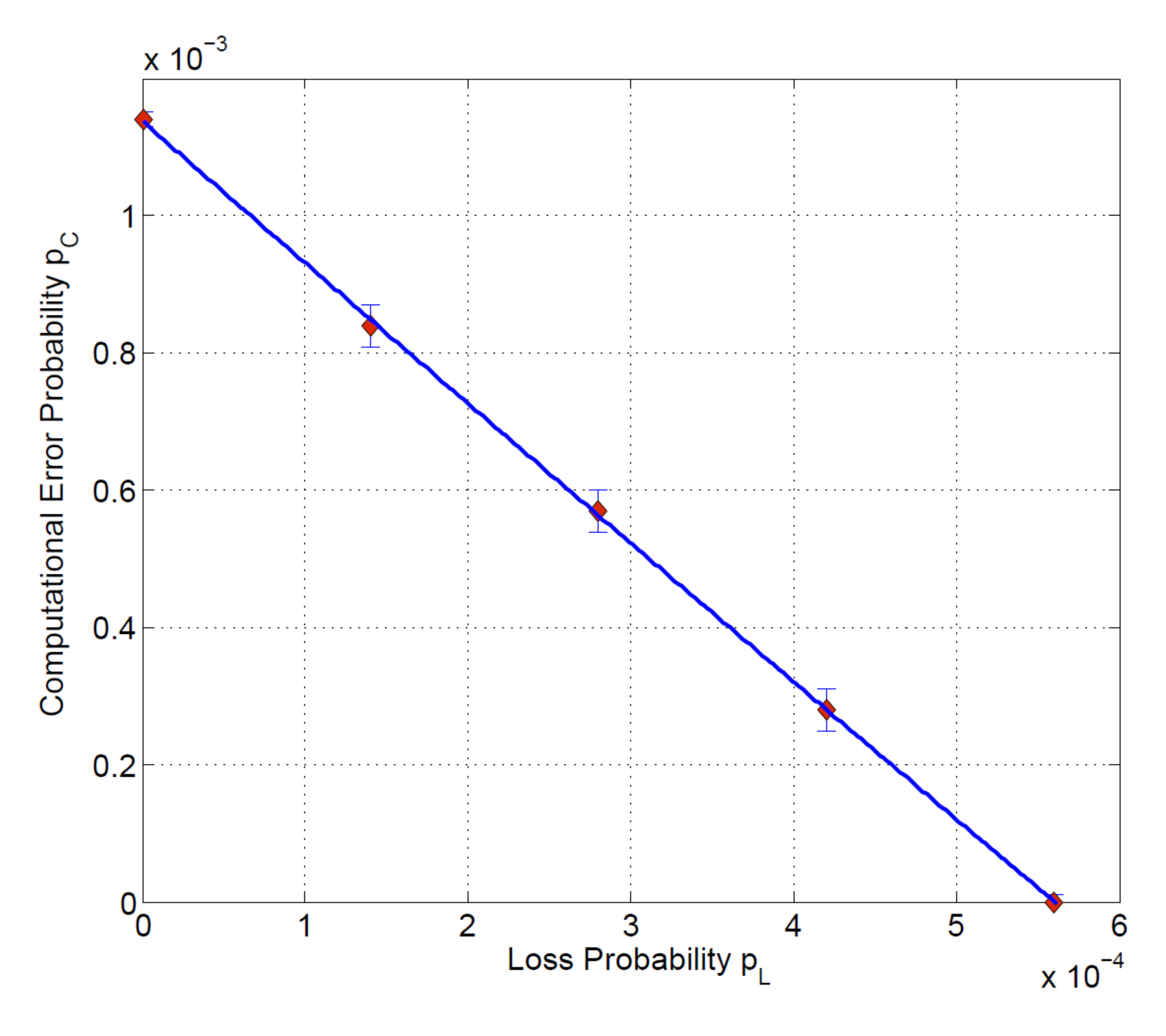}\\
  \caption{Tradeoff curve resulting from a quadratic fit of the calculated thresholds, below which fault-tolerance is achieved.}\label{fig_tradeoff}
\end{figure}

Figure \ref{fig_tradeoff} shows a compromise curve between loss probability and computational error probability. The thresholds showed in fig.~\ref{fit_errors} correspond to the two extremal points in fig.~\ref{fig_tradeoff}. The three points in the middle where each one calculated by choosing a value for $p_L$ and then obtaining a computational threshold as usual. The area underlying the curve is where fault-tolerant quantum computation is possible with our proposed scheme.

\section{Conclusion and Outlook}

We have proposed a particular way of building a fault-tolerant optical computer which using single photon gates and quantum dots. These are thriving technologies which promise scalability and precise control.

The computational error threshold is shown to lie slightly above $10^{-3}$, which is high enough to be compatible with the error rate predicted by \cite{lindner2009}. The efficiencies for detectors and photon emission needed in order to get a connected cluster are, however, extremely high. In future work we will consider a slightly different approach to the error correction which incorporates probabilistic gates into a framework similar to that of \cite{sean2010thresholds}, in the hope of alleviating these stringent requirements. 

The introduction of coupled quantum dots in \cite{economou2010optically} can be exploited to obtain 2-dimensional cluster states, thus reducing our dependence on fusion gates. An open question is whether using this alternative approach would relax the requirements for photon emission efficiency.

\section{Acknowledgements}

The authors would like to acknowledge enlightening conversations with S. Barret, H. Bombin, S. Devitt, S. Economou, R. Raussendorf and T. Stace. Most of the numerical calculations were possible due to the High Performance Cluster of Imperial College. This work is supported by the Engineering and Physical Sciences Research Council.

\bibliography{biblio}

\begin{thebibliography}{10}%
\makeatletter
\providecommand \@ifxundefined [1]{%
 \ifx #1\undefined \expandafter \@firstoftwo
 \else \expandafter \@secondoftwo
\fi
}%
\providecommand \@ifnum [1]{%
 \ifnum #1\expandafter \@firstoftwo
 \else \expandafter \@secondoftwo
\fi
}%
\providecommand \enquote [1]{``#1''}%
\providecommand \bibnamefont  [1]{#1}%
\providecommand \bibfnamefont [1]{#1}%
\providecommand \citenamefont [1]{#1}%
\providecommand\href[0]{\@sanitize\@href}%
\providecommand\@href[1]{\endgroup\@@startlink{#1}\endgroup\@@href}%
\providecommand\@@href[1]{#1\@@endlink}%
\providecommand \@sanitize [0]{\begingroup\catcode`\&12\catcode`\#12\relax}%
\@ifxundefined \pdfoutput {\@firstoftwo}{%
 \@ifnum{\z@=\pdfoutput}{\@firstoftwo}{\@secondoftwo}%
}{%
 \providecommand\@@startlink[1]{\leavevmode\special{html:<a href="#1">}}%
 \providecommand\@@endlink[0]{\special{html:</a>}}%
}{%
 \providecommand\@@startlink[1]{%
  \leavevmode
  \pdfstartlink
   attr{/Border[0 0 1 ]/H/I/C[0 1 1]}%
   user{/Subtype/Link/A<</Type/Action/S/URI/URI(#1)>>}%
  \relax
 }%
 \providecommand\@@endlink[0]{\pdfendlink}%
}%
\providecommand \url  [0]{\begingroup\@sanitize \@url }%
\providecommand \@url [1]{\endgroup\@href {#1}{\urlprefix}}%
\providecommand \urlprefix [0]{URL }%
\providecommand \Eprint[0]{\href }%
\@ifxundefined \urlstyle {%
  \providecommand \doi [1]{doi:\discretionary{}{}{}#1}%
}{%
  \providecommand \doi [0]{doi:\discretionary{}{}{}\begingroup
  \urlstyle{rm}\Url }%
}%
\providecommand \doibase [0]{http://dx.doi.org/}%
\providecommand \Doi[1]{\href{\doibase#1}}%
\providecommand \bibAnnote [3]{%
  \BibitemShut{#1}%
  \begin{quotation}\noindent
    \textsc{Key:}\ #2\\\textsc{Annotation:}\ #3%
  \end{quotation}%
}%
\providecommand \bibAnnoteFile [2]{%
  \IfFileExists{#2}{\bibAnnote {#1} {#2} {\input{#2}}}{}%
}%
\providecommand \typeout [0]{\immediate \write \m@ne }%
\providecommand \selectlanguage [0]{\@gobble}%
\providecommand \bibinfo [0]{\@secondoftwo}%
\providecommand \bibfield [0]{\@secondoftwo}%
\providecommand \translation [1]{[#1]}%
\providecommand \BibitemOpen[0]{}%
\providecommand \bibitemStop [0]{}%
\providecommand \bibitemNoStop [0]{.\EOS\space}%
\providecommand \EOS [0]{\spacefactor3000\relax}%
\providecommand \BibitemShut [1]{\csname bibitem#1\endcsname}%
\bibitem{knill2001scheme}%
  \BibitemOpen
  \bibfield{author}{%
  \bibinfo {author} {\bibfnamefont{E.}~\bibnamefont{Knill}}, \bibinfo {author}
  {\bibfnamefont{R.}~\bibnamefont{Laflamme}},\ and\ \bibinfo {author}
  {\bibfnamefont{G.}~\bibnamefont{Milburn}},\ }%
  \bibfield{journal}{%
  \bibinfo {journal} {Nature}\ }%
  \textbf{\bibinfo {volume} {409}},\ \bibinfo {pages} {46} (\bibinfo {year}
  {2001})%
  \bibAnnoteFile{NoStop}{knill2001scheme}%
\bibitem{dawson2005}%
  \BibitemOpen
  \bibfield{author}{%
  \bibinfo {author} {\bibfnamefont{M.~A.}\ \bibnamefont{Nielsen}}\ and\
  \bibinfo {author} {\bibfnamefont{C.~M.}\ \bibnamefont{Dawson}},\ }%
  \bibfield{journal}{%
  \bibinfo {journal} {Phys. Rev. A}\ }%
  \textbf{\bibinfo {volume} {71}},\ \bibinfo {pages} {042323} (\bibinfo {year}
  {2005})%
  \bibAnnoteFile{NoStop}{dawson2005}%
\bibitem{varnava2007loss}%
  \BibitemOpen
  \bibfield{author}{%
  \bibinfo {author} {\bibfnamefont{M.}~\bibnamefont{Varnava}}, \bibinfo
  {author} {\bibfnamefont{D.}~\bibnamefont{Browne}},\ and\ \bibinfo {author}
  {\bibfnamefont{T.}~\bibnamefont{Rudolph}},\ }%
  \bibfield{journal}{%
  \bibinfo {journal} {New Journal of Physics}\ }%
  \textbf{\bibinfo {volume} {9}},\ \bibinfo {pages} {203} (\bibinfo {year}
  {2007})%
  \bibAnnoteFile{NoStop}{varnava2007loss}%
\bibitem{raussendorf2001one}%
  \BibitemOpen
  \bibfield{author}{%
  \bibinfo {author} {\bibfnamefont{R.}~\bibnamefont{Raussendorf}}\ and\
  \bibinfo {author} {\bibfnamefont{H.}~\bibnamefont{Briegel}},\ }%
  \bibfield{journal}{%
  \bibinfo {journal} {Phys. Rev. Lett.}\ }%
  \textbf{\bibinfo {volume} {86}},\ \bibinfo {pages} {5188} (\bibinfo {year}
  {2001})%
  \bibAnnoteFile{NoStop}{raussendorf2001one}%
\bibitem{raussendorf2006fault}%
  \BibitemOpen
  \bibfield{author}{%
  \bibinfo {author} {\bibfnamefont{R.}~\bibnamefont{Raussendorf}}, \bibinfo
  {author} {\bibfnamefont{J.}~\bibnamefont{Harrington}},\ and\ \bibinfo
  {author} {\bibfnamefont{K.}~\bibnamefont{Goyal}},\ }%
  \bibfield{journal}{%
  \bibinfo {journal} {Ann. Phys.}\ }%
  \textbf{\bibinfo {volume} {321}},\ \bibinfo {pages} {2242} (\bibinfo {year}
  {2006})%
  \bibAnnoteFile{NoStop}{raussendorf2006fault}%
\bibitem{raussendorf2007topological}%
  \BibitemOpen
  \bibfield{author}{%
  \bibinfo {author} {\bibfnamefont{R.}~\bibnamefont{Raussendorf}}, \bibinfo
  {author} {\bibfnamefont{J.}~\bibnamefont{Harrington}},\ and\ \bibinfo
  {author} {\bibfnamefont{K.}~\bibnamefont{Goyal}},\ }%
  \bibfield{journal}{%
  \bibinfo {journal} {N. Jour. Phys.}\ }%
  \textbf{\bibinfo {volume} {9}},\ \bibinfo {pages} {199} (\bibinfo {year}
  {2007})%
  \bibAnnoteFile{NoStop}{raussendorf2007topological}%
\bibitem{nielsen2004optical}%
  \BibitemOpen
  \bibfield{author}{%
  \bibinfo {author} {\bibfnamefont{M.}~\bibnamefont{Nielsen}},\ }%
  \bibfield{journal}{%
  \bibinfo {journal} {Phys. Rev. Let.}\ }%
  \textbf{\bibinfo {volume} {93}},\ \bibinfo {pages} {40503} (\bibinfo {year}
  {2004})%
  \bibAnnoteFile{NoStop}{nielsen2004optical}%
\bibitem{browne2005resource}%
  \BibitemOpen
  \bibfield{author}{%
  \bibinfo {author} {\bibfnamefont{D.}~\bibnamefont{Browne}}\ and\ \bibinfo
  {author} {\bibfnamefont{T.}~\bibnamefont{Rudolph}},\ }%
  \bibfield{journal}{%
  \bibinfo {journal} {Phys. Rev. Lett.}\ }%
  \textbf{\bibinfo {volume} {95}},\ \bibinfo {pages} {10501} (\bibinfo {year}
  {2005})%
  \bibAnnoteFile{NoStop}{browne2005resource}%
\bibitem{lindner2009}%
  \BibitemOpen
  \bibfield{author}{%
  \bibinfo {author} {\bibfnamefont{N.~H.}\ \bibnamefont{Lindner}}\ and\
  \bibinfo {author} {\bibfnamefont{T.}~\bibnamefont{Rudolph}},\ }%
  \bibfield{journal}{%
  \bibinfo {journal} {Phys. Rev. Lett.}\ }%
  \textbf{\bibinfo {volume} {103}},\ \bibinfo {pages} {113602} (\bibinfo
  {month} {Sep}\ \bibinfo {year} {2009})%
  \bibAnnoteFile{NoStop}{lindner2009}%
\bibitem{nielsen-molmer2010}%
  \BibitemOpen
  \bibfield{author}{%
  \bibinfo {author} {\bibfnamefont{A.~E.~B.}\ \bibnamefont{Nielsen}}\ and\
  \bibinfo {author} {\bibfnamefont{K.}~\bibnamefont{M\o{}lmer}},\ }%
  \bibfield{journal}{%
  \bibinfo {journal} {Phys. Rev. A}\ }%
  \textbf{\bibinfo {volume} {81}},\ \bibinfo {pages} {043822} (\bibinfo {year}
  {2010})%
  \bibAnnoteFile{NoStop}{nielsen-molmer2010}%
\bibitem{li2010demand}%
  \BibitemOpen
  \bibfield{author}{%
  \bibinfo {author} {\bibfnamefont{Y.}~\bibnamefont{Li}}, \bibinfo {author}
  {\bibfnamefont{L.}~\bibnamefont{Aolita}},\ and\ \bibinfo {author}
  {\bibfnamefont{L.}~\bibnamefont{Kwek}},\ }%
  \bibfield{journal}{%
  \bibinfo {journal} {Arxiv preprint arXiv:1003.1742}}%
   (\bibinfo {year} {2010})%
  \bibAnnoteFile{NoStop}{li2010demand}%
\bibitem{Raus07}%
  \BibitemOpen
  \bibfield{author}{%
  \bibinfo {author} {\bibfnamefont{R.}~\bibnamefont{Raussendorf}}\ and\
  \bibinfo {author} {\bibfnamefont{J.}~\bibnamefont{Harrington}},\ }%
  \bibfield{journal}{%
  \bibinfo {journal} {Phys. Rev. Lett.}\ }%
  \textbf{\bibinfo {volume} {98}},\ \bibinfo {pages} {190504} (\bibinfo {year}
  {2007}),\ \bibinfo {note} {quant-ph/0610082}%
  \bibAnnoteFile{NoStop}{Raus07}%
\bibitem{Raus07d}%
  \BibitemOpen
  \bibfield{author}{%
  \bibinfo {author} {\bibfnamefont{R.}~\bibnamefont{Raussendorf}}, \bibinfo
  {author} {\bibfnamefont{J.}~\bibnamefont{Harrington}},\ and\ \bibinfo
  {author} {\bibfnamefont{K.}~\bibnamefont{Goyal}},\ }%
  \bibfield{journal}{%
  \bibinfo {journal} {New J. Phys.}\ }%
  \textbf{\bibinfo {volume} {9}},\ \bibinfo {pages} {199} (\bibinfo {year}
  {2007}),\ \bibinfo {note} {quant-ph/0703143}%
  \bibAnnoteFile{NoStop}{Raus07d}%
\bibitem{Brav98}%
  \BibitemOpen
  \bibfield{author}{%
  \bibinfo {author} {\bibfnamefont{S.~B.}\ \bibnamefont{Bravyi}}\ and\ \bibinfo
  {author} {\bibfnamefont{A.~Y.}\ \bibnamefont{Kitaev}},\ }%
  \bibfield{journal}{%
  \bibinfo {journal} {quant-ph/9811052}}%
   (\bibinfo {year} {1998})%
  \bibAnnoteFile{NoStop}{Brav98}%
\bibitem{Denn02}%
  \BibitemOpen
  \bibfield{author}{%
  \bibinfo {author} {\bibfnamefont{E.}~\bibnamefont{Dennis}}, \bibinfo {author}
  {\bibfnamefont{A.}~\bibnamefont{Kitaev}}, \bibinfo {author}
  {\bibfnamefont{A.}~\bibnamefont{Landahl}},\ and\ \bibinfo {author}
  {\bibfnamefont{J.}~\bibnamefont{Preskill}},\ }%
  \bibfield{journal}{%
  \bibinfo {journal} {J. Math. Phys.}\ }%
  \textbf{\bibinfo {volume} {43}},\ \bibinfo {pages} {4452} (\bibinfo {year}
  {2002}),\ \bibinfo {note} {quant-ph/0110143}%
  \bibAnnoteFile{NoStop}{Denn02}%
\bibitem{Fowl08}%
  \BibitemOpen
  \bibfield{author}{%
  \bibinfo {author} {\bibfnamefont{A.~G.}\ \bibnamefont{Fowler}}, \bibinfo
  {author} {\bibfnamefont{A.~M.}\ \bibnamefont{Stephens}},\ and\ \bibinfo
  {author} {\bibfnamefont{P.}~\bibnamefont{Groszkowski}},\ }%
  \bibfield{journal}{%
  \bibinfo {journal} {Phys. Rev. A}\ }%
  \textbf{\bibinfo {volume} {80}},\ \bibinfo {pages} {052312} (\bibinfo {year}
  {2009}),\ \bibinfo {note} {arXiv:0803.0272}%
  \bibAnnoteFile{NoStop}{Fowl08}%
\bibitem{Wang09}%
  \BibitemOpen
  \bibfield{author}{%
  \bibinfo {author} {\bibfnamefont{D.~S.}\ \bibnamefont{Wang}}, \bibinfo
  {author} {\bibfnamefont{A.~G.}\ \bibnamefont{Fowler}}, \bibinfo {author}
  {\bibfnamefont{A.~M.}\ \bibnamefont{Stephens}},\ and\ \bibinfo {author}
  {\bibfnamefont{L.~C.~L.}\ \bibnamefont{Hollenberg}},\ }%
  \bibfield{journal}{%
  \bibinfo {journal} {Quant. Info. Comp.}\ }%
  \textbf{\bibinfo {volume} {10}},\ \bibinfo {pages} {456} (\bibinfo {year}
  {2010}),\ \bibinfo {note} {arXiv:0905.0531}%
  \bibAnnoteFile{NoStop}{Wang09}%
\bibitem{Stoc08}%
  \BibitemOpen
  \bibfield{author}{%
  \bibinfo {author} {\bibfnamefont{R.}~\bibnamefont{Stock}}\ and\ \bibinfo
  {author} {\bibfnamefont{D.~F.~V.}\ \bibnamefont{James}},\ }%
  \bibfield{journal}{%
  \bibinfo {journal} {Phys. Rev. Lett.}\ }%
  \textbf{\bibinfo {volume} {102}},\ \bibinfo {pages} {170501} (\bibinfo {year}
  {2009}),\ \bibinfo {note} {arXiv:0808.1591}%
  \bibAnnoteFile{NoStop}{Stoc08}%
\bibitem{devitt2009architectural}%
  \BibitemOpen
  \bibfield{author}{%
  \bibinfo {author} {\bibfnamefont{S.}~\bibnamefont{Devitt}}, \bibinfo {author}
  {\bibfnamefont{A.}~\bibnamefont{Fowler}}, \bibinfo {author}
  {\bibfnamefont{A.}~\bibnamefont{Stephens}}, \bibinfo {author}
  {\bibfnamefont{A.}~\bibnamefont{Greentree}}, \bibinfo {author}
  {\bibfnamefont{L.}~\bibnamefont{Hollenberg}}, \bibinfo {author}
  {\bibfnamefont{W.}~\bibnamefont{Munro}},\ and\ \bibinfo {author}
  {\bibfnamefont{K.}~\bibnamefont{Nemoto}},\ }%
  \bibfield{journal}{%
  \bibinfo {journal} {N. Jour. Phys.}\ }%
  \textbf{\bibinfo {volume} {11}},\ \bibinfo {pages} {083032} (\bibinfo {year}
  {2009})%
  \bibAnnoteFile{NoStop}{devitt2009architectural}%
\bibitem{VanM09}%
  \BibitemOpen
  \bibfield{author}{%
  \bibinfo {author} {\bibfnamefont{R.}~\bibnamefont{Van{ }Meter}}, \bibinfo
  {author} {\bibfnamefont{T.~D.}\ \bibnamefont{Ladd}}, \bibinfo {author}
  {\bibfnamefont{A.~G.}\ \bibnamefont{Fowler}},\ and\ \bibinfo {author}
  {\bibfnamefont{Y.}~\bibnamefont{Yamamoto}},\ }%
  \bibfield{journal}{%
  \bibinfo {journal} {International Journal of Quantum Information}}%
   (\bibinfo {year} {2010}),\ \bibinfo {note} {to appear; preprint
  arXiv:0906.2686}%
  \bibAnnoteFile{NoStop}{VanM09}%
\bibitem{DiVi09}%
  \BibitemOpen
  \bibfield{author}{%
  \bibinfo {author} {\bibfnamefont{D.~P.}\ \bibnamefont{DiVincenzo}},\ }%
  \bibfield{journal}{%
  \bibinfo {journal} {arXiv:0905.4839}}%
   (\bibinfo {year} {2009}),\ \bibinfo {note} {{N}obel Symposium on Qubits for
  Quantum Information}%
  \bibAnnoteFile{NoStop}{DiVi09}%
\bibitem{Fowl09}%
  \BibitemOpen
  \bibfield{author}{%
  \bibinfo {author} {\bibfnamefont{A.~G.}\ \bibnamefont{Fowler}}\ and\ \bibinfo
  {author} {\bibfnamefont{K.}~\bibnamefont{Goyal}},\ }%
  \bibfield{journal}{%
  \bibinfo {journal} {Quant. Info. Comput.}\ }%
  \textbf{\bibinfo {volume} {9}},\ \bibinfo {pages} {721} (\bibinfo {year}
  {2009}),\ \bibinfo {note} {arXiv:0805.3202}%
  \bibAnnoteFile{NoStop}{Fowl09}%
\bibitem{Note1}%
  \BibitemOpen
  \bibinfo {note} {This amounts to saying that X and Z errors can be considered
  independently.}%
  \bibAnnoteFile{Stop}{Note1}%
\bibitem{bravyi2005universal}%
  \BibitemOpen
  \bibfield{author}{%
  \bibinfo {author} {\bibfnamefont{S.}~\bibnamefont{Bravyi}}\ and\ \bibinfo
  {author} {\bibfnamefont{A.}~\bibnamefont{Kitaev}},\ }%
  \bibfield{journal}{%
  \bibinfo {journal} {Phys. Rev. A}\ }%
  \textbf{\bibinfo {volume} {71}},\ \bibinfo {pages} {22316} (\bibinfo {year}
  {2005})%
  \bibAnnoteFile{NoStop}{bravyi2005universal}%
\bibitem{imamog1999quantum}%
  \BibitemOpen
  \bibfield{author}{%
  \bibinfo {author} {\bibfnamefont{A.}~\bibnamefont{Imamog~Lu}}, \bibinfo
  {author} {\bibfnamefont{D.}~\bibnamefont{Awschalom}}, \bibinfo {author}
  {\bibfnamefont{G.}~\bibnamefont{Burkard}}, \bibinfo {author}
  {\bibfnamefont{D.}~\bibnamefont{DiVincenzo}}, \bibinfo {author}
  {\bibfnamefont{D.}~\bibnamefont{Loss}}, \bibinfo {author}
  {\bibfnamefont{M.}~\bibnamefont{Sherwin}},\ and\ \bibinfo {author}
  {\bibfnamefont{A.}~\bibnamefont{Small}},\ }%
  \bibfield{journal}{%
  \bibinfo {journal} {Physical review letters}\ }%
  \textbf{\bibinfo {volume} {83}},\ \bibinfo {pages} {4204} (\bibinfo {year}
  {1999})%
  \bibAnnoteFile{NoStop}{imamog1999quantum}%
\bibitem{imamoglu2003quantum}%
  \BibitemOpen
  \bibfield{author}{%
  \bibinfo {author} {\bibfnamefont{A.}~\bibnamefont{Imamoglu}},\ }%
  \bibfield{journal}{%
  \bibinfo {journal} {Physica E: Low-dimensional Systems and Nanostructures}\
  }%
  \textbf{\bibinfo {volume} {16}},\ \bibinfo {pages} {47} (\bibinfo {year}
  {2003})%
  \bibAnnoteFile{NoStop}{imamoglu2003quantum}%
\bibitem{loss2000electron}%
  \BibitemOpen
  \bibfield{author}{%
  \bibinfo {author} {\bibfnamefont{D.}~\bibnamefont{Loss}}, \bibinfo {author}
  {\bibfnamefont{G.}~\bibnamefont{Burkard}},\ and\ \bibinfo {author}
  {\bibfnamefont{D.}~\bibnamefont{DiVincenzo}},\ }%
  \bibfield{journal}{%
  \bibinfo {journal} {Journal of Nanoparticle Research}\ }%
  \textbf{\bibinfo {volume} {2}},\ \bibinfo {pages} {401} (\bibinfo {year}
  {2000})%
  \bibAnnoteFile{NoStop}{loss2000electron}%
\bibitem{Note2}%
  \BibitemOpen
  \bibinfo {note} {We can equally consider a rotation of $2\pi $, since
  $R_{Y}(2\pi ) \equiv -I$, since this can also be used to obtain RE qubits up
  to phases.}%
  \bibAnnoteFile{Stop}{Note2}%
\bibitem{stace2009thresholds}%
  \BibitemOpen
  \bibfield{author}{%
  \bibinfo {author} {\bibfnamefont{T.}~\bibnamefont{Stace}}, \bibinfo {author}
  {\bibfnamefont{S.}~\bibnamefont{Barrett}},\ and\ \bibinfo {author}
  {\bibfnamefont{A.}~\bibnamefont{Doherty}},\ }%
  \bibfield{journal}{%
  \bibinfo {journal} {Phys.l Rev. Lett.}\ }%
  \textbf{\bibinfo {volume} {102}},\ \bibinfo {pages} {200501} (\bibinfo {year}
  {2009})%
  \bibAnnoteFile{NoStop}{stace2009thresholds}%
\bibitem{stace2010error}%
  \BibitemOpen
  \bibfield{author}{%
  \bibinfo {author} {\bibfnamefont{T.}~\bibnamefont{Stace}}\ and\ \bibinfo
  {author} {\bibfnamefont{S.}~\bibnamefont{Barrett}},\ }%
  \bibfield{journal}{%
  \bibinfo {journal} {Phys. Rev. A}\ }%
  \textbf{\bibinfo {volume} {81}},\ \bibinfo {pages} {22317} (\bibinfo {year}
  {2010})%
  \bibAnnoteFile{NoStop}{stace2010error}%
\bibitem{Note3}%
  \BibitemOpen
  \bibinfo {note} {The simulations we present here were done using the Imperial
  College High Performance Cluster. Preliminary estimates (100 simulations per
  point) were taken running our algorithm on an Intel Q8200 processor at 2.33
  GHz and 3.5 Gb of RAM, for up to $d\leq 13$.}%
  \bibAnnoteFile{Stop}{Note3}%
\bibitem{sean2010thresholds}%
  \BibitemOpen
  \bibfield{author}{%
  \bibinfo {author} {\bibfnamefont{S.~D.}\ \bibnamefont{Barrett}}\ and\
  \bibinfo {author} {\bibfnamefont{T.}~\bibnamefont{Stace}},\ }%
  \bibfield{journal}{%
  \bibinfo {journal} {Arxiv preprint arXiv:1005.2456}}%
   (\bibinfo {year} {2010})%
  \bibAnnoteFile{NoStop}{sean2010thresholds}%
\bibitem{economou2010optically}%
  \BibitemOpen
  \bibfield{author}{%
  \bibinfo {author} {\bibfnamefont{S.}~\bibnamefont{Economou}}, \bibinfo
  {author} {\bibfnamefont{N.}~\bibnamefont{Lindner}},\ and\ \bibinfo {author}
  {\bibfnamefont{T.}~\bibnamefont{Rudolph}},\ }%
  \bibfield{journal}{%
  \bibinfo {journal} {Arxiv preprint arXiv:1003.2410}}%
   (\bibinfo {year} {2010})%
  \bibAnnoteFile{NoStop}{economou2010optically}%
\end{thebibliography}%

\end{document}